\documentstyle[aps,12pt]{revtex}

\def\>{\rangle}

\title{Entanglement Properties of Some Fractional Quantum Hall Liquids}
\author{Bei Zeng$^{1}$, Hui Zhai$^{2}$ and Zhan Xu$^{1,2}$}
\address{1. Department of Physics, Tsinghua University, \\Beijing, 100084, China
        \\ 2. Center for Advanced Study, Tsinghua University, \\Beijing, 100084, China
        }
\begin{document}
\maketitle

\begin{abstract}
    We study the entanglement properties of some fractional quantum Hall liquids. We
    calculate the entanglement of the Laughlin wave function and
    the wave functions that are generated by the K-matrix using the
    modified entanglement measure of indistinguishable fermions that is
    first proposed by Pa\v{s}kauskas and You\cite{py}.

\end{abstract}
\pacs{03.67.-a,73.43.-f}
\date{}

\section{Introduction}
    Entanglement is no doubt an intriguing property of composite quantum systems. It is a correlation that is stronger
    than any classical correlation. It is found that entanglement is not only of interests in the interpretation
    of the foundation of quantum mechanics,  but also a resource useful in quantum information processing and quantum
    computation. Although the theory of entanglement is widely developed in the systems of distinguishable particles,
    only very recently the entanglement properties in identical particle systems began to attract much
    attention\cite{slm}\cite{sckll}\cite{lbll}\cite{gf}\cite{esbl} in the fields of quantum information
    and quantum computation.

    Quite recently,  it is realized that applying the theory of entanglement developed in quantum information science
    to the field of condensed matter physics may give us new insight in these problems, especially the abundant
    entanglement properties of ground state wave function and the wave functions related to the quantum phase transitions\cite{on}\cite{oafr}.
    Some systems have already been extensively studied,  such as the models of Heissenberg spin chain\cite{wz}
    and harmonic chain\cite{aepw}. It should be emphasized that these systems are all related to the crystal lattice,
    thus the entanglement can be calculated using the side entropy suggested by Zanardi\cite{zn} or just using the
    measure of entanglement suggested by the entanglement theory of distinguishable particle systems by taking account of the
    fact that each site can be viewed as one "party" in quantum information science.

    It is noticed that the fractional quantum Hall liquid is a kind of strongly correlated
    quantum fluid,  in which quantum correlation plays an essential role\cite{prange}.
    As the external magnetic field perpendicular to the two-dimensional election gas increases,  the Hall conductance as
    well as the filling factor jumps from one value to another. Correspondingly,  a
    change of the wave function of the system takes place. It is well-known that the fractional quantum Hall
    effect
    with the filling fractions $\nu=1/\mbox{(odd integer)}$ has been explained by Laughlin\cite{laughlin}.
    Besides these states,  the quantum Hall liquid possesses an extremely rich internal structure,  which is classified in
    terms of so-called K-matrix\cite{wena}. So far it is clear that
    the fractional quantum Hall system experiences quantum phase
    transition,  and some characters of one fractional quantum Hall state
    are essentially different from those of another state. Seeking some order or index to
    describe these essential characters can facilitate further
    understanding of the quantum phase transition. Many work has been done
    along this direction\cite{wenb}. Recently it is noticed that the
    states in  fractional quantum Hall liquid have very sophisticated entanglement properties
    \cite{yu},  and different states may have different entanglement
    properties. Thus it attracts our attention to
    apply the theory of entanglement developed in quantum information science
    to investigate the entanglement properties in this system.

    However,  since there is no longer any site in the liquid,  at first sight it seems that we cannot
    use the entanglement measure of
    distinguishable particle systems to calculate the entanglement of this system. Surprisingly, it is found
    by Pa\v{s}kauskas and You\cite{py} that the von Neumann entropy of the reduced single particle density
    matrix remains to be a good
    entanglement measure for two identical particles, which is a natural extension of the entanglement measure of
    distinguishable particles.Thus,  this entanglement
    measure of identical particles is sufficient to meet our need here.

    In this paper we study the entanglement properties of the Laughlin wave function and
    the wave functions that are generated by the K-matrix of quantum Hall liquid.
    First in section 2 we will explain the meaning of the entanglement measure of indistinguishable fermions that is proposed
    by Pa\v{s}kauskas and
    You,  and further point out the relationship between this measure and the measure used in the distinguishable particle
    systems. We will find that a slight modification is needed in Pa\v{s}kauskas and
    You's measure. Then we calculate the entanglement of the Laughlin wave
    function and the wave functions that are generated by the K-matrix in the simplest case that the particle number $N$
    is $2$ in section 2 using the slightly modified entanglement measure of indistinguishable fermions. In section 3
    the more sophisticated case with the particle number lager than
    $2$ is considered.Some discussions about our results will be given in Section 4.

\section{Entanglement Measure of Indistinguishable Fermions}
    The first entanglement measure of indistinguishable fermions is  introduced
    by J.Schliemann et. al\cite{slm}\cite{lbll}.In \cite{slm}, they claimed that the concept of separablility
    of a state in composite systems of
    fermions should be defined in terms of  that the state can be expressed in the form of a single Slater
    determinant. In \cite{lbll},  they find that the wave function of two fermions can be written in the following standard form

    \begin{equation}
    {|\Psi\rangle=\frac{1}{\sqrt{\sum_{i=1}^{k}|z_{i}|^{2}}}\sum\limits_{i=1}^{k}z_{i}f^{+}_{a_{1}(i)}f^{+}_{a_{2}(i)}|0\rangle}
    \end{equation}

    where $f^{+}_{a_{1}(i)}|0\rangle$ and $f^{+}_{a_{2}(i)}|0\rangle$ represent the orthonormal basis of single
    particle space. Following this idea,  they defined an entanglement measure
    of two fermions as follows

    \begin{equation}
    \eta(|\Psi\rangle)=|\langle\bar{\Psi}|\Psi\rangle|
    \end{equation}

    where $|\bar{\Psi}\rangle$ is the dual state of $|\Psi\rangle$.It can be verified that $\eta(|\Psi\rangle)=0$
    if and only if $|\Psi\rangle$ can be expressed in the form of a single Slater
    determinant and $\eta(|\Psi\rangle)$ is a smooth function ranging from  $\eta(|\Psi\rangle)=0$ for the separable state
    to $\eta(|\Psi\rangle)=1$ for the  maximum  correlated state.

    Later, Pa\v{s}kauskas and You\cite{py} suggested another entanglement measure for two
    fermions.They found that the von Neumann entropy of the reduced single particle density matrix remains to be a good
    entanglement measure for two identical particles.Thus,  for the
    state (1), the entanglement can be calculated as

    \begin{equation}
    S_{f}^{PY}=-tr[\rho^{f}ln\rho^{f}]=-1n2-4\sum\limits_{k=1}^{\leq{n/2}}|z_{k}/2|^{2}ln(|z_{k}/2|^{2})
    \end{equation}
    where $n$ is the dimension of the single particle space.
     $S_{f}$ is also a smooth function ranging from  $S_{f}=ln2$ for the separable state to  $S_{f}=lnn_{e}$ for the  maximum correlated
    state, where $n_{e}$ is the maximal even number less or equal to $n$.

    However,  a problem with this measure still remains, for the un-correlated two-fermion state gives $S_{f}=ln2$
    rather than $0$. We point out here that this problem can be easily understood because it is noticed that the extra $ln2$
    is just the entanglement contained in the antisymmetry of the wave function in the first quantized representation of two
    identical fermions. It is already known that this extra entanglement is of no use in quantum information processing. Since in
    identical particle systems it is important to find the entanglement beyond that involved in the (anti)symmetry induced by quantum
    statistics,  we need to get rid of this extra $ln2$, i.e. this
    entanglement measure needs a slight modification

    \begin{equation}
    S_{f}=-tr[\rho^{f}ln\rho^{f}]-ln2=-21n2-4\sum\limits_{k=1}^{\leq{n/2}}|z_{k}/2|^{2}ln(|z_{k}/2|^{2})
    \end{equation}

    It must be emphasized that although this modification is quite slight,  it is very important in the sense
    that this modification enables us to obtain the same value of entanglement measure when the system of identical
    particles is reduced to a system of distinguishable particles.
    To be more concrete,  consider the following state of two qubits written in the form of
    the Schmidt decomposition

    \begin{equation}
    |\psi\rangle=\alpha|00\rangle+\beta|11\rangle
    \end{equation}
    whose second quantized counterpart is
    \begin{equation}
    |\psi\rangle=(\alpha a^{\dag}b^{\dag}+\beta c^{\dag}d^{\dag})|0\rangle
    \end{equation}
    where $|\alpha|^{2}|+|\beta|^{2}=1$.
    \\It is easy to get
    \begin{equation}
    S_{f}(|\psi\rangle)=-|\alpha|^{2}ln|\alpha|^{2}-|\beta|^{2}ln|\beta|^{2}
    \end{equation}

    This is in accordance with the von Neumann entropy of two-qubit
    state(5). It is also a very easy task to show that when the two fermions are distinguishable,
    what the
    modified entanglement measure(4) give us is just the von Neumann
    entropy. In this sense we can say the modified entanglement
    measure(4) is the measure suitable for two-fermion systems,  either
    identical or
    distinguishable.

    It is also noticed that to calculate entanglement of two fermions with eq(4) has another three advantages. First,  we can see from eq(4) the close relationship between this measure and the measure
    used in the distinguishable particle systems.  Second,  we can either calculate this
    measure directly from the first quantized representation using the skills developed to calculate the
    entanglement measure of distinguishable particles or from the second quantized
    representation by just calculating the one particle density matrix and
    diagnosing
    it.  Third,  this measure can be extended to multi-particle case naturally to calculate the entanglement between
    one-particle
    and the other particles in the system with a slight modification

    \begin{equation}
    S'_{f}=-tr[\rho^{f}ln\rho^{f}]-lnN
    \end{equation}
    where $N$ is the particle number. It is noticed that the entanglement measure(2) cannot be directly
    extended to multi-particle case\cite{esbl}.

    Although in multi-particle case this measure cannot give us all the entanglement properties of the state,  we can get at
    least some information about the entanglement properties of the state.  In fact it is well-known that in both the
    distinguishable and identical particle case the question of quantify the multi-particle entanglement is still open.

    Considered all in all,  we will use eq(4) and eq(5) to explore the entanglement properties in quantum Hall liquid.

\section{Entanglement in Quantum Hall Liquid ($N=2$ case)}
    The Laughlin wave function for a quantum Hall liquid with filling factor $1/m$ reads

    \begin{equation}
    \psi_{m}(z_{1}, . . . , z_{N})=\prod\limits_{j<k}(z_{j}-z_{k})^{m}exp\left(-\frac{1}{4}\sum\limits_{i}|z_{i}|^{2}\right)
    \end{equation}
    where $z_{i}$s are complex coordinates of the electrons and $m$ is a positive odd number.
    It is already known that the case $m=1$ corresponds to the integer quantum Hall effect, where the Laughlin wave
    function is just a single Slater determinant,  i. e.

    \begin{equation}
    \psi_{1}(z_{1}, . . . , z_{N})=exp\left(-\frac{1}{4}\sum\limits_{i}|z_{i}|^{2}\right)\times
    det\left(
    \begin{array}{cccc}
    1, &z_{1}, &. . . , &z_{1}^{N}\\
    . . . &. . . &. . . &. . . \\
    1, &z_{N}, &. . . , &z_{N}^{N}
    \end{array}
    \right)
    \end{equation}
    thus $\psi_{1}$ is separable.

    However,  when $m\neq1$, $\psi_{m}$ cannot be expressed in the form of a single Slater determinant,
     i.e. it is
    entangled. In order to explore the entanglement properties of Laughlin wave
    function in the case $m>1$,  we need to calculate the amount of entanglement contained in these functions.  First,  we consider
    the simplest case that the particle number $N=2$,  then the Laughlin wave function will be

    \begin{eqnarray}
    \psi_{m}(z_{1}, z_{2})&=&(z_{1}-z_{2})^{m}exp\left(-\frac{1}{4}(|z_{1}|^{2}+|z_{2}|^{2})\right)\nonumber\\
                        &=&\left(\sum\limits_{k=0}^{m}(-1)^{k}C_{m}^{k}z_{1}^{m-k}z_{2}^{k}\right)exp\left(-\frac{1}{4}(|z_{1}|^{2}+|z_{2}|^{2})\right)\nonumber\\
                        &=&\left(\sum\limits_{k=0}^{(m-1)/2}(-1)^{k}C_{m}^{k}(z_{1}^{m-k}z_{2}^{k}-z_{1}^{k}z_{2}^{m-k})\right)exp\left(-\frac{1}{4}(|z_{1}|^{2}+|z_{2}|^{2})\right)\nonumber\\
                        &=&\left(\sum\limits_{k=0}^{(m-1)/2}(-1)^{k}C_{m}^{k}\times det\left(
    \begin{array}{cc}
    z_{1}^{m-k}, &z_{1}^{k}\\
    z_{2}^{m-k}, &z_{2}^{k}
    \end{array}
    \right)\right)exp\left(-\frac{1}{4}(|z_{1}|^{2}+|z_{2}|^{2})\right)\nonumber\\
    \end{eqnarray}
    \\Obviously,
    to calculate the measure of entanglement(4) from the first quantization point of view is a very difficult
    task,  for this is a problem concerning
    continuous variable entanglement which always cannot be
    obtained
    analytically.  Then we need to tackle this problem from the
    second quantization point of view. Fortunately,  if the one particle space is chosen
    appropriately, the calculation can be done within a finite
    dimensional space.It is noticed that these functions in the lowest Landau level with different angular momentum are
    written as:

    \begin{equation}
    \{f_{i}(z)=A_{i}z^{i}exp(-\frac{1}{4}|z|^{2})\}_{i=0}^{m}
    \end{equation}
    The family of functions $f_{i}(z), i=0. . . m$ form an orthogonal basis of one particle space, where $A_{i}$
    is the normalization factor, i.e.

    \begin{equation}
    A_{i}=\frac{1}{\sqrt{\int\limits_{x=-\infty}^{\infty}\int\limits_{y=-\infty}^{\infty}(x^{2}+y^{2})^{i}
    exp(-\frac{1}{2}(x^{2}+y^{2}))}}=\frac{1}{\sqrt{\pi2^{i+1}i!}}
    \end{equation}
    and $z=x+iy$.

    Define a set of creation operators
    $\{a_{i}^{\dag}\}_{i=0}^{m}$ corresponding to $f_i(z)$ by

    \begin{equation}
    \langle{z}|a_{i}^{\dag}|0\rangle=f_{i}(z)
    \end{equation}
    \\Thus we can rewrite $\psi_{m}$ in the second quantized form as

    \begin{eqnarray}
    |\psi_{m}\rangle&=&\left(\frac{1}{\sqrt{\sum\limits_{k=0}^{(m-1)/2}\frac{(C_{m}^{k})^{2}}{A_{k}^{2}A_{m-k}^{2}}}}\right)
                        \sum\limits_{k=0}^{(m-1)/2}\left[\frac{C_{m}^{k}}{A_{k}A_{m-k}}\right]a_{m-k}^{\dag}a_{k}^{\dag}|0\rangle\nonumber\\
                    &=&2^{-\frac{m-1}{2}}\sum\limits_{k=0}^{(m-1)/2}\sqrt{C_{m}^{k}}a_{m-k}^{\dag}a_{k}^{\dag}|0\rangle
    \end{eqnarray}

    It is easy to find that $|\psi_{m}\rangle$ in eq(12) has the form of the Slater decomposition (1),  thus we can calculate
    the entanglement of $|\psi_{m}\rangle$ directly using formula(4):

    \begin{eqnarray}
    S_{f}(|\psi_{m}\rangle)&=&-2ln2-2^{(m-1)}\sum\limits_{k=0}^{(m-1)/2}C_{m}^{k}ln\left(2^{-(m+1)}C_{m}^{k}\right)\nonumber\\
                            &=&-2ln2-2^{-m}\sum\limits_{k=0}^{m}ln\left(2^{-(m+1)}C_{m}^{k}\right)
    \end{eqnarray}

    And the variation of $S_{f}(|\psi_{m}\rangle)$ with $t=(m-1)/2$ is shown in Figure 1.
    We can see from Figure 1 that the entanglement increases with $m$ increasing.

    We know that a quasihole excitation above $|\psi _{m}\rangle $ is
    described
    by $\psi \left( \xi \right) =\sqrt{N\left( \xi , \xi ^{\ast }\right) }
    \prod\limits_{i}\left( \xi -z_{i}\right) \psi _{m}$.  These quasiholes can form new quantum fluid in the second level,  leading
    to more complicated filling fraction,  such as $\nu =\frac{7}{2}$.
    It is well-known that there are hierarchical fractional quantum Hall states. The wave functions of such states can be constructed with
    the help of K-matrix\cite{wena}. As an
    example, the state characterized  by using
    $K=\left(
    \begin{array}{cc}
    m & 1 \\
    1 & -2
    \end{array}
    \right) $ has a filling fraction $\nu =\frac{1}{m-\left(- \frac{
    1}{2}\right) }=\frac{2}{2m+1}$, and is described by the generalized
    Laughlin wave
    function:

    \begin{eqnarray}
    \phi_{m}(z_{1}, z_{2})=(z_{1}-z_{2})^{m}\int\int
    d\xi_{1}d\xi_{2}(\xi_{1}-z_{1})(\xi_{1}-z_{2})(\xi_{2}-z_{1})(\xi_{2}-z_{2})\nonumber\\
    \times (\xi_{1}^{\ast}-\xi_{1}^{\ast})^{2}exp\left(-\frac{1}{3}\left(|\xi_{1}|^{2}+|\xi_{2}|^{2}\right)\right)
    exp\left(-\frac{1}{4}\left(|z_{1}|^{2}+|z_{2}|^{2}\right)\right)
    \end{eqnarray}

    Physically,  these states have two quasiholes above the
    $|\psi_{m}\rangle $ state,  and then these quasiholes form quantum Hall liquid with $
    \nu =\frac{1}{2}$.  Integrating over the coordinates of the
    quasiholes,  one obtains a new dancing pattern of these electrons
    which is different from that of $|\psi _{m}\rangle $.  Naturally,
    it leads to a change of the entanglement property.  It is found that

    \begin{eqnarray}
    \int\int d\xi_{1}d\xi_{2}(\xi_{1}-z_{1})(\xi_{1}-z_{2})(\xi_{2}-z_{1})(\xi_{2}-z_{2})(\xi_{1}^{\ast}-\xi_{1}^{\ast})^{2}
    \nonumber\\
    \times exp\left(-\frac{1}{3}\left(|\xi_{1}|^{2}+|\xi_{2}|^{2}\right)\right)=-162\pi(z_{1}^{2}+z_{2}^{2})
    \end{eqnarray}

    Thus $\phi_{m}(z_{1}, z_{2})$ can be rewritten as:

    \begin{equation}
    \phi_{m}(z_{1}, z_{2})=(z_{1}-z_{2})^{m}(z_{1}^{2}+z_{2}^{2})
    exp\left(-\frac{1}{4}\left(|z_{1}|^{2}+|z_{2}|^{2}\right)\right)
    \end{equation}

    And in the second quantized form the normalized
    state vector of $\phi_{m}(z_{1}, z_{2})$ reads:

    \begin{equation}
    |\phi_{m}\rangle=\left(\frac{1}{\sqrt{\sum\limits_{k=0}^{(m+1)/2}\frac{(C_{m}^{k}+C_{m}^{k-2})^{2}}{A_{k}^{2}A_{m-k+2}^{2}}}}\right)
                        \sum\limits_{k=0}^{(m+1)/2}\left[\frac{C_{m}^{k}+C_{m}^{k-2}}{A_{k}A_{m-k+2}}\right]a_{m-k+2}^{\dag}a_{k}^{\dag}|0\rangle
    \end{equation}
    \\Obviously,  $|\phi_{m}\rangle$ in eq(17) also has the form of the Slater decomposition(1),  thus we can calculate
    the entanglement of $|\psi_{m}\rangle$ directly using formula(4):

    \begin{eqnarray}
    S_{f}(|\phi_{m}\rangle)&=&-4\sum\limits_{k=0}^{(m+1)/2}
    \left(\frac{\frac{(C_{m}^{k}+C_{m}^{k-2})^{2}}{A_{k}^{2}A_{m-k+2}^{2}}}{2\left(\sum\limits_{k=0}^{(m+1)/2}\frac{(C_{m}^{k}+C_{m}^{k-2})^{2}}{A_{k}^{2}A_{m-k+2}^{2}}\right)}\right)\nonumber\\
    &&\times ln\left(\frac{\frac{(C_{m}^{k}+C_{m}^{k-2})^{2}}{A_{k}^{2}A_{m-k+2}^{2}}}{2\left(\sum\limits_{k=0}^{(m+1)/2}\frac{(C_{m}^{k}+C_{m}^{k-2})^{2}}{A_{k}^{2}A_{m-k+2}^{2}}\right)}\right)
    -2ln2
    \end{eqnarray}

    And the variation of $S_{f}(|\psi_{m}\rangle)$ with $t=(m-1)/2$ is shown in Figure 1. We can see from Figure 1 that the entanglement also increases when $m$ is
    increasing. Comparing the two curves in Figure 1 we can find that
    for each
    $m$, $S_{f}(|\psi_{m}\rangle)>S_{f}(|\phi_{m}\rangle)$. This
    fact can be understood by noting that adding a
    hierarchical structure will result in the increase of
    quantum entanglement. It is interesting to notice that $S_{f}(|\psi_{3}\rangle)=S_{f}(|\phi_{1}\rangle)$. This can
    also be found from their explicit expression, i. e.

    \begin{equation}
    |\psi_{3}\rangle=(z_{1}-z_{2})^{3}\sim(a_{0}^{\dag}a_{3}^{\dag}+\sqrt{3}a_{1}^{\dag}a_{2}^{\dag})
    \end{equation}

    \begin{equation}
    |\phi_{1}\rangle=(z_{1}-z_{2})(z_{1}^{2}+z_{2}^{2})\sim(\sqrt{3}a_{0}^{\dag}a_{3}^{\dag}+a_{1}^{\dag}a_{2}^{\dag})
    \end{equation}

\section{Entanglement in Quantum Hall Liquid ($N>2$ case)}
    Now we turn to the much more difficult case $N>2$.  As mentioned above,  we can only get some information of the entanglement
    between one particle and other particles using eq(5). We again use second quantized representation to calculate
    the entanglement of this special kind of multi-particle states.

    Take $N=3$ Laughlin wave function for example

    \begin{eqnarray}
    \psi'_{m}(z_{1}, z_{2}, z_{3})=(z_{1}-z_{2})^{m}(z_{1}-z_{3})^{m}(z_{2}-z_{3})^{m}\nonumber\\
    \times exp\left(-\frac{1}{4}\left(|z_{1}|^{2}+|z_{2}|^{2}+|z_{3}|^{2}\right)\right)
    \end{eqnarray}
    We have

    \begin{equation}
    S'_{f}(\psi'_{m})=-tr[\rho^{f}ln\rho^{f}]-ln3
    \end{equation}
    The second quantized form of $|\psi'_{3}\rangle$ is

    \begin{eqnarray}
    |\psi'_{3}\rangle=(\frac{1}{A_{0}A_{3}A_{6}}a_{0}^{\dag}a_{3}^{\dag}a_{6}^{\dag}
    +\frac{3}{A_{0}A_{4}A_{5}}a_{0}^{\dag}a_{4}^{\dag}a_{5}^{\dag}
    +\frac{6}{A_{1}A_{3}A_{5}}a_{1}^{\dag}a_{3}^{\dag}a_{5}^{\dag}\nonumber\\
    +\frac{3}{A_{1}A_{2}A_{6}}a_{1}^{\dag}a_{2}^{\dag}a_{6}^{\dag}
    +\frac{15}{A_{2}A_{3}A_{4}}a_{2}^{\dag}a_{3}^{\dag}a_{4}^{\dag})|0\rangle
    \end{eqnarray}

    It is noted for $|\psi'_{3}\rangle$,  the single-particle  density
    matrix(up to a normalization factor)
    $\rho_{\mu\nu}^{f}=\langle\psi'_{3}|a_{\mu}^{\dag}a_{\nu}|\psi'_{3}\rangle$
    has no off-diagonal elements,  thus $S'_{f}(|\psi'_{3}\rangle)$ can easily be
    calculated from the form of $|\psi'_{3}\rangle$.

    In fact,  since $\psi'_{m}(z_{1}, z_{2}, z_{3})$ is a homogeneous polynomial of $z_{1}$, $z_{2}$and$z_{3}$ apart from an exponential factor,  there
    will be no off-diagonal elements of the single-particle density matrix, for arbitrary positive odd value of $m$. This is also
    true for any particle number $N$. This fact makes the calculation of $S'_{f}$ much easier. Our result for the $N=3$ case
    is shown in Figure 2. It can be seen from Figure 2 that
    $S'_{f}$also increases when $m$ is increasing.

    Similar things happen in the calculation of $S'_{f}$ of a state that is generated by the same kind of
    K-matrix mentioned above,  since in these
    wave functions the polynomial is also homogeneous. Take $N=3$
    for example

    \begin{eqnarray}
    \phi'_{m}(z_{1}, z_{2}, z_{3})=(z_{1}-z_{2})^{m}(z_{1}-z_{3})^{m}(z_{2}-z_{3})^{m}\nonumber\\
    \times exp\left(-\frac{1}{4}\left(|z_{1}|^{2}+|z_{2}|^{2}+|z_{3}|^{2}\right)\right)\nonumber\\
    \times\int\int
    d\xi_{1}d\xi_{2}(\xi_{1}-z_{1})(\xi_{1}-z_{2})(\xi_{1}-z_{3})(\xi_{2}-z_{1})\nonumber\\
    \times(\xi_{2}-z_{2})(\xi_{2}-z_{3})(\xi_{1}^{\ast}-\xi_{1}^{\ast})^{2}
    exp\left(-\frac{1}{3}\left(|\xi_{1}|^{2}+|\xi_{2}|^{2}\right)\right)
    \end{eqnarray}
    Since

    \begin{eqnarray}
    \int\int
    d\xi_{1}d\xi_{2}(\xi_{1}-z_{1})(\xi_{1}-z_{2})(\xi_{1}-z_{3})(\xi_{2}-z_{1})(\xi_{2}-z_{2})(\xi_{2}-z_{3})\nonumber\\
    \times(\xi_{1}^{\ast}-\xi_{1}^{\ast})^{2}
    exp\left(-\frac{1}{3}\left(|\xi_{1}|^{2}+|\xi_{2}|^{2}\right)\right)=-162\pi(z_{1}^{2}z_{2}^2+z_{1}^{2}z_{3}^2+z_{2}^{2}z_{3}^2)
    \end{eqnarray}
    We get

    \begin{eqnarray}
    \phi'_{m}(z_{1}, z_{2}, z_{3})=(z_{1}-z_{2})^{m}(z_{1}-z_{3})^{m}(z_{2}-z_{3})^{m}
    (z_{1}^{2}z_{2}^2+z_{1}^{2}z_{3}^2+z_{2}^{2}z_{3}^2)\nonumber\\
    \times exp\left(-\frac{1}{4}\left(|z_{1}|^{2}+|z_{2}|^{2}\right)\right)
    \end{eqnarray}
    which is also a homogeneous polynomial of $z_{1}$, $z_{2}$and$z_{3}$ apart from an exponential factor.

    Thus the calculation of $S'_{f}$ of this kind of wave functions is also quite
    easy. Our result for $N=3$ case is shown in Figure 2. It can be seen from Figure 2 that
    $S'_{f}$also increases when $m$ is increasing. However,  we can
    see that at this time $S_{f}(|\psi'_{3}\rangle)$ is no longer equal
    to $S_{f}(|\phi'_{1}\rangle)$. Comparing the two curves in Figure 2 we can find that
    for each
    $m$, the relation $S_{f}(|\psi'_{m}\rangle)>S_{f}(|\phi'_{m}\rangle)$ still holds.

    The results of $N=2$ and $N=3$ cases for the Laughlin wave
    function are shown in Figure 3. It can be seen that the amount
    of entanglement between one electron and the other electrons
    increases with the electron number N. And the K-matrix case is
    shown in Figure 4. However, this case is quite sophisticated since
    the amount
    of entanglement between one electron and the other electrons
    only increases with the electron number N besides the case
    $m=3$.

    Now we turn to another kind of dancing patterns of these strongly correlated electrons.
    We have already shown that the $\nu =1$ state is a separable state,
    while if there exists some quasihole excitation above it,  and then
    these quasiholes form a Laughlin-type state with filling
    fraction
    $\nu =\frac{1}{m-1}$ ,  where $m$ is an odd integer.  It will
    result in entanglement between electrons.  The K-matrix is written
    as $\left(
    \begin{array}{cc}
    1 & 1 \\
    1 & -\left( m-1\right)
    \end{array}
    \right) $ , and the corresponding filling fractions are $\nu=\frac{1}{1-\frac{1}{m-1}}$.
    Thus the  wave functions associated with this K-matrix take the form(take the case $N=4$ for example):

    \begin{eqnarray}
    \chi_{m}(z_{1}, z_{2}, z_{3}, z_{4})=(z_{1}-z_{2})(z_{1}-z_{3})(z_{2}-z_{3})(z_{1}-z_{4})(z_{2}-z_{4})(z_{3}-z_{4})\nonumber\\
    \times exp\left(-\frac{1}{4}\left(|z_{1}|^{2}+|z_{2}|^{2}+|z_{3}|^{2}+|z_{4}|^{2}\right)\right)\nonumber\\
    \times\int\int
    d\xi_{1}d\xi_{2}(\xi_{1}-z_{1})(\xi_{1}-z_{2})(\xi_{1}-z_{3})(\xi_{1}-z_{4})(\xi_{2}-z_{1})(\xi_{2}-z_{2})\nonumber\\
    \times(\xi_{2}-z_{3})(\xi_{2}-z_{4})(\xi_{1}^{\ast}-\xi_{2}^{\ast})^{(m-1)}
    exp\left(-\frac{1}{3}\left(|\xi_{1}|^{2}+|\xi_{2}|^{2}\right)\right)
    \end{eqnarray}

    Using the method above we can also get some information of the
    entanglement properties these wave functions.
    For $S'_{f}(\chi_{m}(z_{1}, z_{2}, z_{3}, z_{4}))$,  our result is shown in Figure 5.
    It is similar to that of Figure 1 ,
    this is consistent with the particle-hole duality picture in
    quantum Hall liquid.  The eventually vanishing of $S^{\prime}_{f}$
    is due to the effect of finite number of electrons.  It is easy to
    find that when $m>2N+1$

    \begin{equation}
    \int\int d\xi_{1}d\xi_{2}\prod\limits_{i=1}^{N}(\xi_{1}-z_{i})(\xi_{2}-z_{i})(\xi_{1}^{\ast}-\xi_{2}^{\ast})^{(m-1)}
    exp\left(-\frac{1}{3}\left(|\xi_{1}|^{2}+|\xi_{2}|^{2}\right)\right)=0
    \end{equation}
    This fact means that the $\nu=1$ $N$-electron system cannot support a Laughlin type
    wave function of quasiholes when $m$ is larger than $2N+1$.

\section{Discussions}
    We calculated the entanglement of the Laughlin wave function and
    the wave functions that are generated by the K-matrix using the
    modified entanglement measure of indistinguishable fermions that is
    first proposed by Pa\v{s}kauskas and You\cite{py} through the second quantized
    approach in this paper.  It is noticed that the fractional quantum Hall effect occurs when the external
    magnetic field is sufficiently high,  the degeneracy of the lowest
    Landau level is large enough that the problem can be treated in
    the subspace of lowest Landau level.  This leads to the fact that the wave
    function can be expressed in an elegant form which is an
    analytical homogeneous polynomial of its arguments apart from a
    Gaussian factor,  and results in a clean second quantized form
    without any off-diagonal elements.  This property enables us to
    write down the entanglement measure in an analytical way.

    Our result shows that for both kinds of wave functions,  the
    amount of entanglement contained in each kind of wave
    function increases with the increase of the parameter
    $m$.
    However,  since it is well-known that the filling fraction
    changes dramatically with the increase of external magnetic
    field $B$,  the entanglement properties is indeed very
    sophisticated in the system of quantum Hall liquid. It is expected that our
    results and methods can shed light on further studies of
    quantum orders in quantum Hall liquid and other physical
    systems.

\section*{ACKNOWLEDGEMENT}
    The authors would like to thank Mr. Jingen Xiang for helpful
    calculations and Prof. Lee Chang for useful discussions. This
    work is supported by National Natural Science Foundation of
    China (Grant No. 90103004).

\section*{Figure captions}

FIG. 1.  The $N=2$ case: The variations of
$S_{f}(|\psi_{m}\rangle)$ and $S_{f}(|\phi_{m}\rangle)$ with
$t=(m-1)/2$ in the units of $ln2$ $bits$. The boxes represent the
value of $S_{f}(|\psi_{m}\rangle)$ of the Laughlin wave functions
and the crosses represent the value of $S_{f}(|\phi_{m}\rangle)$
of the wave functions generated by the K-matrix.

FIG. 2. The $N=3$ case: The variations of
$S_{f}(|\psi'_{m}\rangle)$ and $S_{f}(|\phi'_{m}\rangle)$ with
$t=(m-1)/2$ in the units of $ln2$ $bits$. The boxes represent the
value of $S_{f}(|\psi'_{m}\rangle)$ of the Laughlin wave functions
and the crosses represent the value of $S_{f}(|\phi'_{m}\rangle)$
of the wave functions generated by the K-matrix..

FIG. 3. The variation of $S_{f}$ of the Laughlin wave functions
with $t=(m-1)/2$ in the units of $ln2$ $bits$. The boxes represent
the value of $S_{f}(|\psi_{m}\rangle)$ of the $N=2$ case and the
crosses represent the value of $S_{f}(|\psi'_{m}\rangle)$ of the
$N=3$ case.

FIG. 4. The variation of $S_{f}$ of the wave functions generated
by the ($m=3$) K-matrix with $t=(m-1)/2$ in the units of $ln2$
$bits$. The boxes represent the value of $S_{f}(|\phi_{m}\rangle)$
of the $N=2$ case and the crosses represent the value of
$S_{f}(|\phi'_{m}\rangle)$ of the $N=3$ case.

FIG. 5. The $N=4$ case: The variation of $S_{f}$ of the wave
function generated by the K-matrix in the units of $ln2$ $bits$.

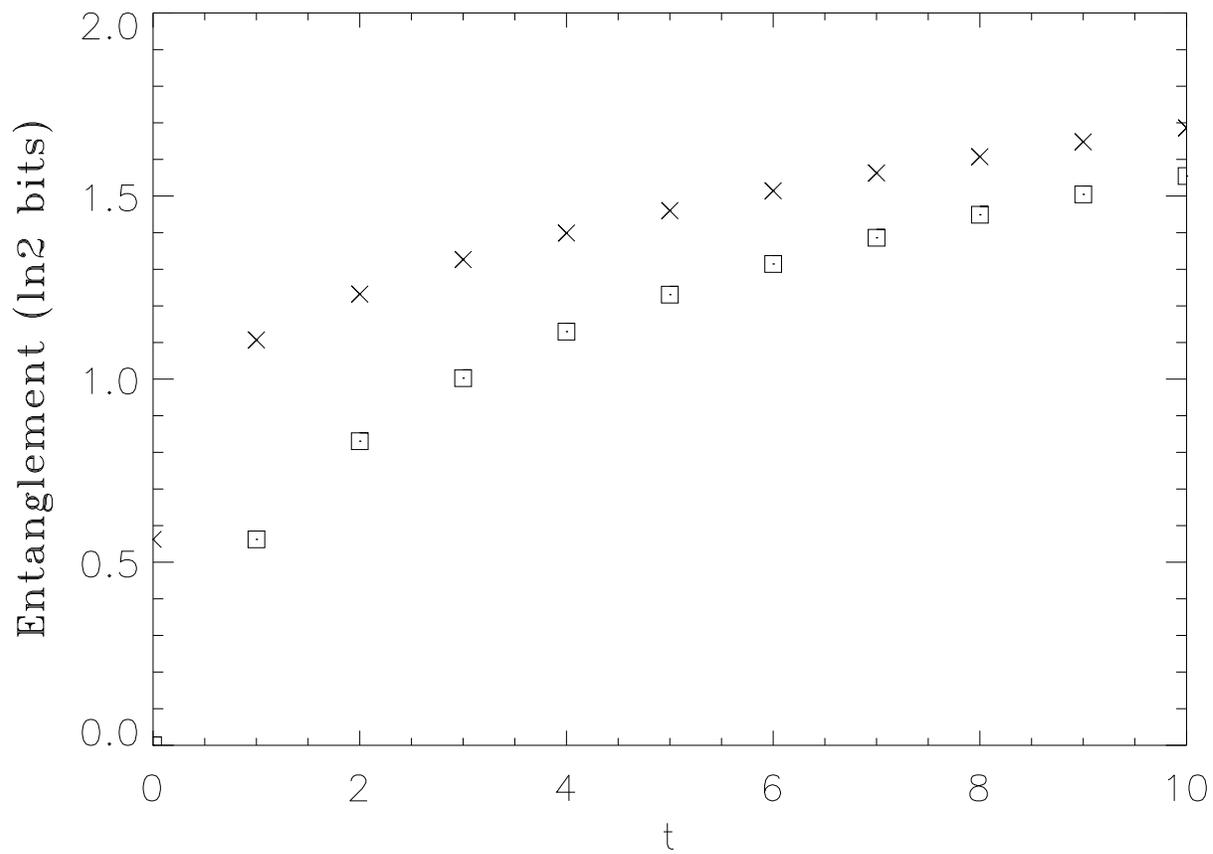
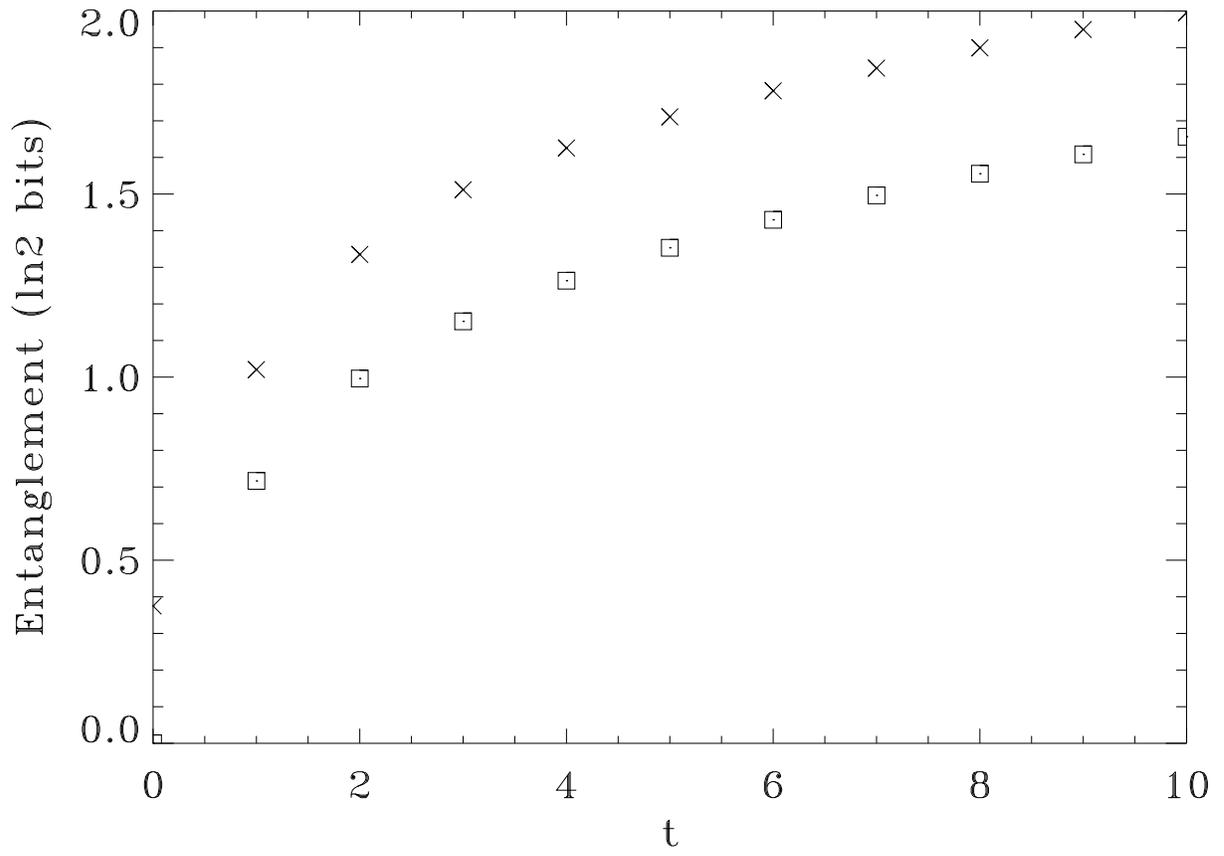
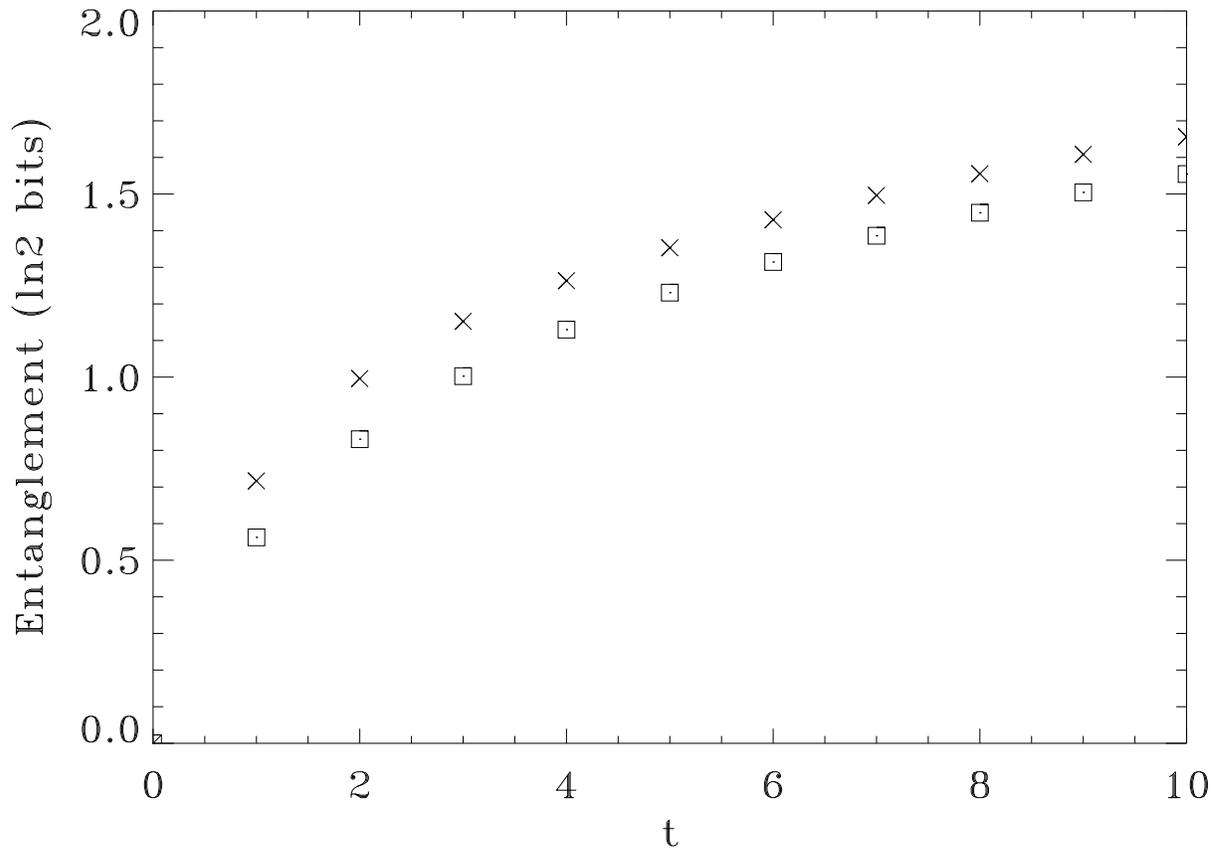
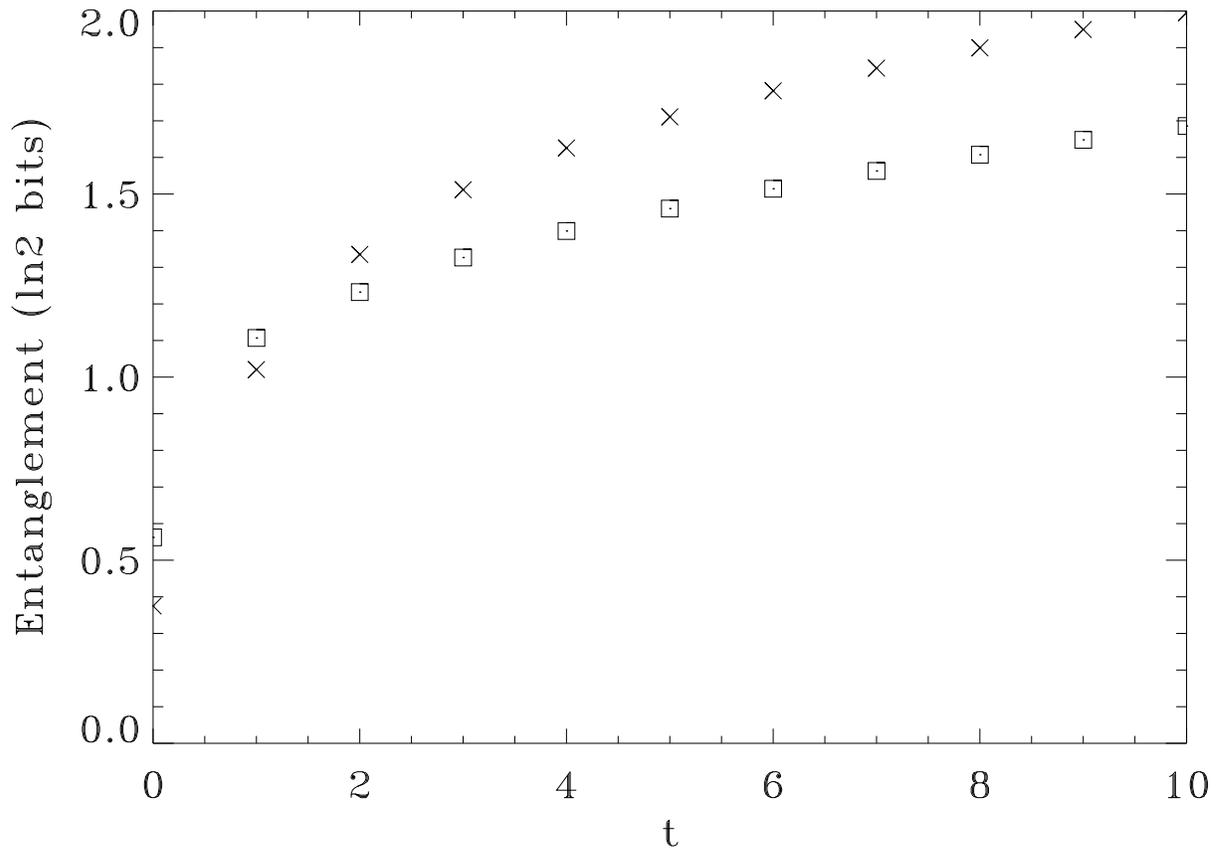
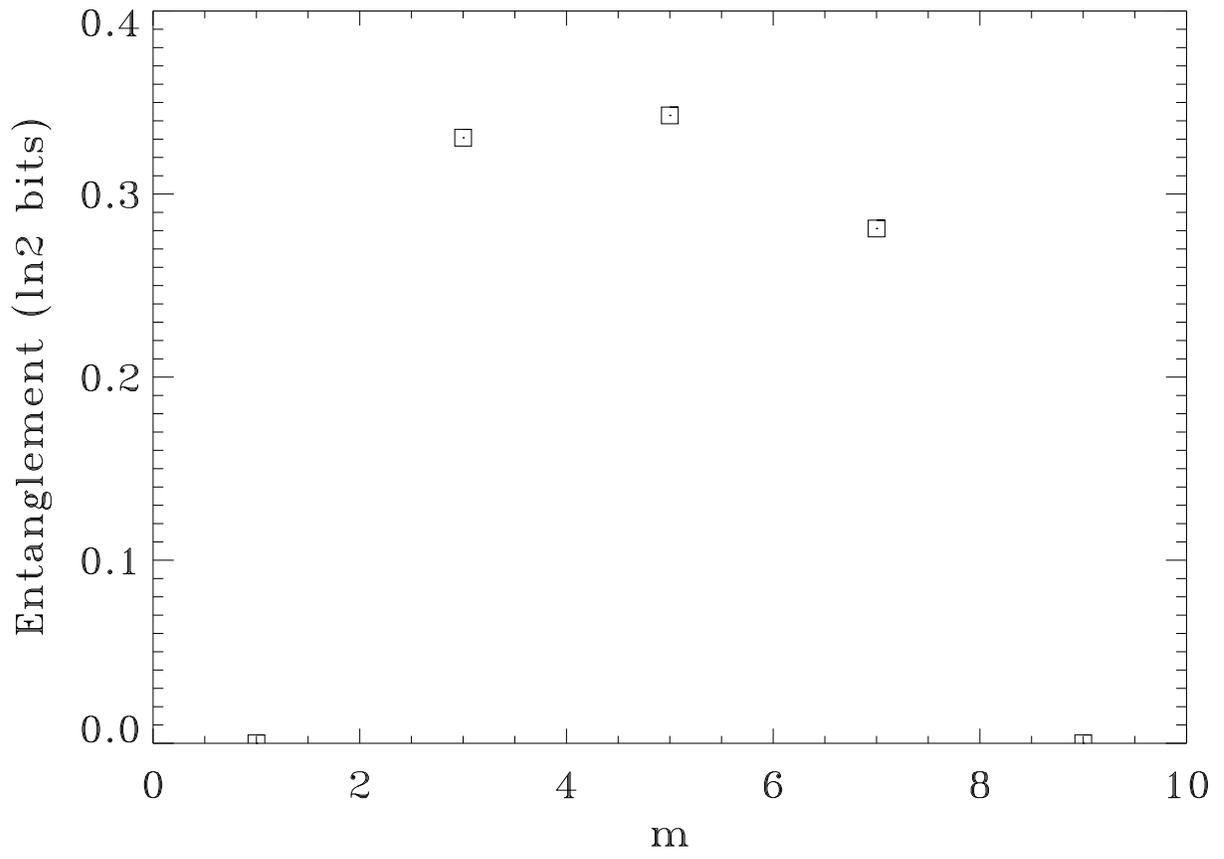
\end{document}